
%
\expandafter\ifx\csname phyzzx\endcsname\relax
 \message{It is better to use PHYZZX format than to
          \string\input\space PHYZZX}\else
 \wlog{PHYZZX macros are already loaded and are not
          \string\input\space again}%
 \endinput \fi
\catcode`\@=11 
\let\rel@x=\relax
\let\n@expand=\relax
\def\pr@tect{\let\n@expand=\noexpand}
\let\protect=\pr@tect
\let\gl@bal=\global
%
%
%
\newfam\cpfam
\newdimen\b@gheight             \b@gheight=12pt
\newcount\f@ntkey               \f@ntkey=0
\def\f@m{\afterassignment\samef@nt\f@ntkey=}
\def\samef@nt{\fam=\f@ntkey \the\textfont\f@ntkey\rel@x}
\def\setstr@t{\setbox\strutbox=\hbox{\vrule height 0.85\b@gheight
                                depth 0.35\b@gheight width\z@ }}
%
%
%
%
%

\font\fourteenrm  =cmr10 scaled\magstep2
\font\twelverm    =cmr12
\font\ninerm      =cmr9
\font\sixrm       =cmr6

\font\fourteenbf  =cmbx10 scaled\magstep2
\font\twelvebf    =cmbx12
\font\ninebf      =cmbx9
\font\sixbf       =cmbx6
\font\seventeeni  =cmmi10 scaled\magstep3    \skewchar\seventeeni='177
\font\fourteeni   =cmmi10 scaled\magstep2     \skewchar\fourteeni='177
\font\twelvei     =cmmi12                       \skewchar\twelvei='177
\font\ninei       =cmmi9                          \skewchar\ninei='177
\font\sixi        =cmmi6                           \skewchar\sixi='177
\font\seventeensy =cmsy10 scaled\magstep3    \skewchar\seventeensy='60
\font\fourteensy  =cmsy10 scaled\magstep2     \skewchar\fourteensy='60
\font\twelvesy    =cmsy10 scaled\magstep1       \skewchar\twelvesy='60
\font\ninesy      =cmsy9                          \skewchar\ninesy='60
\font\sixsy       =cmsy6                           \skewchar\sixsy='60

\font\fourteenex  =cmex10 scaled\magstep2
\font\twelveex    =cmex10 scaled\magstep1

\font\fourteensl  =cmsl10 scaled\magstep2
\font\twelvesl    =cmsl12
\font\ninesl      =cmsl9

\font\fourteenit  =cmti10 scaled\magstep2
\font\twelveit    =cmti12
\font\nineit      =cmti9
\font\fourteentt  =cmtt10 scaled\magstep2
\font\twelvett    =cmtt12
\font\fourteencp  =cmcsc10 scaled\magstep2
\font\twelvecp    =cmcsc10 scaled\magstep1
\font\tencp       =cmcsc10
%
%
\def\fourteenf@nts{\relax
    \textfont0=\fourteenrm          \scriptfont0=\tenrm
      \scriptscriptfont0=\sevenrm
    \textfont1=\fourteeni           \scriptfont1=\teni
      \scriptscriptfont1=\seveni
    \textfont2=\fourteensy          \scriptfont2=\tensy
      \scriptscriptfont2=\sevensy
    \textfont3=\fourteenex          \scriptfont3=\twelveex
      \scriptscriptfont3=\tenex
    \textfont\itfam=\fourteenit     \scriptfont\itfam=\tenit
    \textfont\slfam=\fourteensl     \scriptfont\slfam=\tensl
    \textfont\bffam=\fourteenbf     \scriptfont\bffam=\tenbf
      \scriptscriptfont\bffam=\sevenbf
    \textfont\ttfam=\fourteentt
    \textfont\cpfam=\fourteencp }
\def\twelvef@nts{\relax
    \textfont0=\twelverm          \scriptfont0=\ninerm
      \scriptscriptfont0=\sixrm
    \textfont1=\twelvei           \scriptfont1=\ninei
      \scriptscriptfont1=\sixi
    \textfont2=\twelvesy           \scriptfont2=\ninesy
      \scriptscriptfont2=\sixsy
    \textfont3=\twelveex          \scriptfont3=\tenex
      \scriptscriptfont3=\tenex
    \textfont\itfam=\twelveit     \scriptfont\itfam=\nineit
    \textfont\slfam=\twelvesl     \scriptfont\slfam=\ninesl
    \textfont\bffam=\twelvebf     \scriptfont\bffam=\ninebf
      \scriptscriptfont\bffam=\sixbf
    \textfont\ttfam=\twelvett
    \textfont\cpfam=\twelvecp }
\def\tenf@nts{\relax
    \textfont0=\tenrm          \scriptfont0=\sevenrm
      \scriptscriptfont0=\fiverm
    \textfont1=\teni           \scriptfont1=\seveni
      \scriptscriptfont1=\fivei
    \textfont2=\tensy          \scriptfont2=\sevensy
      \scriptscriptfont2=\fivesy
    \textfont3=\tenex          \scriptfont3=\tenex
      \scriptscriptfont3=\tenex
    \textfont\itfam=\tenit     \scriptfont\itfam=\seveni  
    \textfont\slfam=\tensl     \scriptfont\slfam=\sevenrm 
    \textfont\bffam=\tenbf     \scriptfont\bffam=\sevenbf
      \scriptscriptfont\bffam=\fivebf
    \textfont\ttfam=\tentt
    \textfont\cpfam=\tencp }
%
%

%
\def\rm{\n@expand\f@m0 }
\def\mit{\n@expand\f@m1 }         
\def\cal{\n@expand\f@m2 }
\def\it{\n@expand\f@m\itfam}
\def\sl{\n@expand\f@m\slfam}
\def\bf{\n@expand\f@m\bffam}
\def\tt{\n@expand\f@m\ttfam}
\def\caps{\n@expand\f@m\cpfam}    
\def\em@{\rel@x\ifnum\f@ntkey=0 \it \else
        \ifnum\f@ntkey=\bffam \it \else \rm \fi \fi }
\def\em{\n@expand\em@}
\def\fourteenpoint{\fourteenf@nts \samef@nt \b@gheight=14pt \setstr@t }
\def\twelvepoint{\twelvef@nts \samef@nt \b@gheight=12pt \setstr@t }
\def\tenpoint{\tenf@nts \samef@nt \b@gheight=10pt \setstr@t }
\normalbaselineskip = 19.2pt plus 0.2pt minus 0.1pt 
\normallineskip = 1.5pt plus 0.1pt minus 0.1pt
\normallineskiplimit = 1.5pt
\newskip\normaldisplayskip
\normaldisplayskip = 14.4pt plus 3.6pt minus 10.0pt 
\newskip\normaldispshortskip
\normaldispshortskip = 6pt plus 5pt
\newskip\normalparskip
\normalparskip = 6pt plus 2pt minus 1pt
\newskip\skipregister
\skipregister = 5pt plus 2pt minus 1.5pt
\newif\ifsingl@
\newif\ifdoubl@
\newif\iftwelv@  \twelv@true
\def\singlespace{\singl@true\doubl@false\spaces@t}
\def\doublespace{\singl@false\doubl@true\spaces@t}
\def\normalspace{\singl@false\doubl@false\spaces@t}
\def\Tenpoint{\tenpoint\twelv@false\spaces@t}
\def\Twelvepoint{\twelvepoint\twelv@true\spaces@t}
\def\spaces@t{\rel@x
      \iftwelv@ \ifsingl@\subspaces@t3:4;\else\subspaces@t1:1;\fi
       \else \ifsingl@\subspaces@t3:5;\else\subspaces@t4:5;\fi \fi
      \ifdoubl@ \multiply\baselineskip by 5
         \divide\baselineskip by 4 \fi }
\def\subspaces@t#1:#2;{
      \baselineskip = \normalbaselineskip
      \multiply\baselineskip by #1 \divide\baselineskip by #2
      \lineskip = \normallineskip
      \multiply\lineskip by #1 \divide\lineskip by #2
      \lineskiplimit = \normallineskiplimit
      \multiply\lineskiplimit by #1 \divide\lineskiplimit by #2
      \parskip = \normalparskip
      \multiply\parskip by #1 \divide\parskip by #2
      \abovedisplayskip = \normaldisplayskip
      \multiply\abovedisplayskip by #1 \divide\abovedisplayskip by #2
      \belowdisplayskip = \abovedisplayskip
      \abovedisplayshortskip = \normaldispshortskip
      \multiply\abovedisplayshortskip by #1
        \divide\abovedisplayshortskip by #2
      \belowdisplayshortskip = \abovedisplayshortskip
      \advance\belowdisplayshortskip by \belowdisplayskip
      \divide\belowdisplayshortskip by 2
      \smallskipamount = \skipregister
      \multiply\smallskipamount by #1 \divide\smallskipamount by #2
      \medskipamount = \smallskipamount \multiply\medskipamount by 2
      \bigskipamount = \smallskipamount \multiply\bigskipamount by 4 }
\def\normalbaselines{ \baselineskip=\normalbaselineskip
   \lineskip=\normallineskip \lineskiplimit=\normallineskip
   \iftwelv@\else \multiply\baselineskip by 4 \divide\baselineskip by 5
     \multiply\lineskiplimit by 4 \divide\lineskiplimit by 5
     \multiply\lineskip by 4 \divide\lineskip by 5 \fi }
\Twelvepoint  
\interlinepenalty=50
\interfootnotelinepenalty=5000
\predisplaypenalty=9000
\postdisplaypenalty=500
\hfuzz=1pt
\vfuzz=0.2pt
\newdimen\HOFFSET  \HOFFSET=0pt
\newdimen\VOFFSET  \VOFFSET=0pt
\newdimen\HSWING   \HSWING=0pt
\dimen\footins=8in
%
%
%
\newskip\pagebottomfiller
\pagebottomfiller=\z@ plus \z@ minus \z@
\def\pagecontents{
   \ifvoid\topins\else\unvbox\topins\vskip\skip\topins\fi
   \dimen@ = \dp255 \unvbox255
   \vskip\pagebottomfiller
   \ifvoid\footins\else\vskip\skip\footins\footrule\unvbox\footins\fi
   \ifr@ggedbottom \kern-\dimen@ \vfil \fi }
\def\makeheadline{\vbox to 0pt{ \skip@=\topskip
      \advance\skip@ by -12pt \advance\skip@ by -2\normalbaselineskip
      \vskip\skip@ \line{\vbox to 12pt{}\the\headline} \vss
      }\nointerlineskip}
\def\makefootline{\baselineskip = 1.5\normalbaselineskip
                 \line{\the\footline}}
\newif\iffrontpage
\newif\ifp@genum
\def\nopagenumbers{\p@genumfalse}
\def\pagenumbers{\p@genumtrue}
\pagenumbers
\newtoks\paperheadline
\newtoks\paperfootline
\newtoks\letterheadline
\newtoks\letterfootline
\newtoks\letterinfo
\newtoks\date
\paperheadline={\hfil}
\paperfootline={\hss\iffrontpage\else\ifp@genum\tenrm\folio\hss\fi\fi}
\letterheadline{\iffrontpage \hfil \else
    \rm \ifp@genum page~~\folio\fi \hfil\the\date \fi}
\letterfootline={\iffrontpage\the\letterinfo\else\hfil\fi}
\letterinfo={\hfil}
\def\monthname{\rel@x\ifcase\month 0/\or January\or February\or
   March\or April\or May\or June\or July\or August\or September\or
   October\or November\or December\else\number\month/\fi}
\def\today{\monthname~\number\day, \number\year}
\date={\today}
\headline=\paperheadline 
\footline=\paperfootline 
\countdef\pageno=1      \countdef\pagen@=0
\countdef\pagenumber=1  \pagenumber=1
\def\advancepageno{\gl@bal\advance\pagen@ by 1
   \ifnum\pagenumber<0 \gl@bal\advance\pagenumber by -1
    \else\gl@bal\advance\pagenumber by 1 \fi
    \gl@bal\frontpagefalse  \swing@ }
\def\folio{\ifnum\pagenumber<0 \romannumeral-\pagenumber
           \else \number\pagenumber \fi }
\def\swing@{\ifodd\pagenumber \gl@bal\advance\hoffset by -\HSWING
             \else \gl@bal\advance\hoffset by \HSWING \fi }
\def\footrule{\dimen@=\prevdepth\nointerlineskip
   \vbox to 0pt{\vskip -0.25\baselineskip \hrule width 0.35\hsize \vss}
   \prevdepth=\dimen@ }
\let\footnotespecial=\rel@x
\newdimen\footindent
\footindent=24pt
\def\Textindent#1{\noindent\llap{#1\enspace}\ignorespaces}
\def\Vfootnote#1{\insert\footins\bgroup
   \interlinepenalty=\interfootnotelinepenalty \floatingpenalty=20000
   \singl@true\doubl@false\Tenpoint
   \splittopskip=\ht\strutbox \boxmaxdepth=\dp\strutbox
   \leftskip=\footindent \rightskip=\z@skip
   \parindent=0.5\footindent \parfillskip=0pt plus 1fil
   \spaceskip=\z@skip \xspaceskip=\z@skip \footnotespecial
   \Textindent{#1}\footstrut\futurelet\next\fo@t}

\def\vfootnote#1{\Vfootnote{${#1}$}}
\def\footnote#1{\attach{#1}\vfootnote{#1}}

\let\footsymbol=\star
\newcount\lastf@@t           \lastf@@t=-1
\newcount\footsymbolcount    \footsymbolcount=0
\newif\ifPhysRev
\def\bumpfootsymbolcount{\rel@x
   \iffrontpage \bumpfootsymbolpos \else \advance\lastf@@t by 1
     \ifPhysRev \bumpfootsymbolneg \else \bumpfootsymbolpos \fi \fi
   \gl@bal\lastf@@t=\pagen@ }
\def\bumpfootsymbolpos{\ifnum\footsymbolcount <0
                            \gl@bal\footsymbolcount =0 \fi
    \ifnum\lastf@@t<\pagen@ \gl@bal\footsymbolcount=0
     \else \gl@bal\advance\footsymbolcount by 1 \fi }
\def\bumpfootsymbolneg{\ifnum\footsymbolcount >0
             \gl@bal\footsymbolcount =0 \fi
         \gl@bal\advance\footsymbolcount by -1 }
\def\fd@f#1 {\xdef\footsymbol{\mathchar"#1 }}
\def\generatefootsymbol{\ifcase\footsymbolcount \fd@f 13F \or \fd@f 279
        \or \fd@f 27A \or \fd@f 278 \or \fd@f 27B \else
        \ifnum\footsymbolcount <0 \fd@f{023 \number-\footsymbolcount }
         \else \fd@f 203 {\loop \ifnum\footsymbolcount >5
                \fd@f{203 \footsymbol } \advance\footsymbolcount by -1
                \repeat }\fi \fi }

\def\nonfrenchspacing{\sfcode`\.=3001 \sfcode`\!=3000 \sfcode`\?=3000
        \sfcode`\:=2000 \sfcode`\;=1500 \sfcode`\,=1251 }
\nonfrenchspacing
\newdimen\d@twidth
{\setbox0=\hbox{s.} \gl@bal\d@twidth=\wd0 \setbox0=\hbox{s}
        \gl@bal\advance\d@twidth by -\wd0 }
\def\removehglue{\loop \unskip \ifdim\lastskip >\z@ \repeat }
\def\roll@ver#1{\removehglue \nobreak \count255 =\spacefactor \dimen@=\z@
        \ifnum\count255 =3001 \dimen@=\d@twidth \fi
        \ifnum\count255 =1251 \dimen@=\d@twidth \fi
    \iftwelv@ \kern-\dimen@ \else \kern-0.83\dimen@ \fi
   #1\spacefactor=\count255 }
\def\step@ver#1{\rel@x \ifmmode #1\else \ifhmode
        \roll@ver{${}#1$}\else {\setbox0=\hbox{${}#1$}}\fi\fi }
\def\attach#1{\step@ver{\strut^{\mkern 2mu #1} }}
%
%
%
\newcount\chapternumber      \chapternumber=0
\newcount\sectionnumber      \sectionnumber=0
\newcount\equanumber         \equanumber=0
\let\chapterlabel=\rel@x
\let\sectionlabel=\rel@x
\newtoks\chapterstyle        \chapterstyle={\Number}
\newtoks\sectionstyle        \sectionstyle={\chapterlabel.\Number}
\newskip\chapterskip         \chapterskip=\bigskipamount
\newskip\sectionskip         \sectionskip=\medskipamount
\newskip\headskip            \headskip=8pt plus 3pt minus 3pt
\newdimen\chapterminspace    \chapterminspace=15pc
\newdimen\sectionminspace    \sectionminspace=10pc
\newdimen\referenceminspace  \referenceminspace=20pc
\def\chapterreset{\gl@bal\advance\chapternumber by 1
   \ifnum\equanumber<0 \else\gl@bal\equanumber=0\fi
   \sectionnumber=0 \let\sectionlabel=\rel@x
   {\pr@tect\xdef\chapterlabel{\the\chapterstyle{\the\chapternumber}}}}
\def\alphabetic#1{\count255='140 \advance\count255 by #1\char\count255}
\def\Alphabetic#1{\count255='100 \advance\count255 by #1\char\count255}
\def\Roman#1{\uppercase\expandafter{\romannumeral #1}}
\def\roman#1{\romannumeral #1}
\def\Number#1{\number #1}
\def\BLANC#1{}
\def\titleparagraphs{\interlinepenalty=9999
     \leftskip=0.03\hsize plus 0.22\hsize minus 0.03\hsize
     \rightskip=\leftskip \parfillskip=0pt
     \hyphenpenalty=9000 \exhyphenpenalty=9000
     \tolerance=9999 \pretolerance=9000
     \spaceskip=0.333em \xspaceskip=0.5em }
\def\titlestyle#1{\par\begingroup \titleparagraphs
     \iftwelv@\fourteenpoint\else\twelvepoint\fi
   \noindent #1\par\endgroup }
\def\spacecheck#1{\dimen@=\pagegoal\advance\dimen@ by -\pagetotal
   \ifdim\dimen@<#1 \ifdim\dimen@>0pt \vfil\break \fi\fi}
\def\chapter#1{\par \penalty-300 \vskip\chapterskip
   \spacecheck\chapterminspace
   \chapterreset \titlestyle{\chapterlabel.~#1}
   \nobreak\vskip\headskip \penalty 30000
   {\pr@tect\wlog{\string\chapter\space \chapterlabel}} }

\def\section#1{\par \ifnum\the\lastpenalty=30000\else
   \penalty-200\vskip\sectionskip \spacecheck\sectionminspace\fi
   \gl@bal\advance\sectionnumber by 1
   {\pr@tect
   \xdef\sectionlabel{\the\sectionstyle\the\sectionnumber}
   \wlog{\string\section\space \sectionlabel}}
   \noindent {\caps\enspace\sectionlabel.~~#1}\par
   \nobreak\vskip\headskip \penalty 30000 }
\def\subsection#1{\par
   \ifnum\the\lastpenalty=30000\else \penalty-100\smallskip \fi
   \noindent\undertext{#1}\enspace \vadjust{\penalty5000}}

\def\undertext#1{\vtop{\hbox{#1}\kern 1pt \hrule}}
\def\APPENDIX#1#2{\par\penalty-300\vskip\chapterskip
   \spacecheck\chapterminspace \chapterreset \xdef\chapterlabel{#1}
   \titlestyle{APPENDIX #2} \nobreak\vskip\headskip \penalty 30000
   \wlog{\string\Appendix~\chapterlabel} }
\def\Appendix#1{\APPENDIX{#1}{#1}}
\def\appendix{\APPENDIX{A}{}}
\def\unnumberedchapters{\let\makechapterlabel=\rel@x
      \let\chapterlabel=\rel@x  \sectionstyle={\BLANC}
      \let\sectionlabel=\rel@x \sequentialequations }
%
%
%
\def\eqname#1{\rel@x {\pr@tect
  \ifnum\equanumber<0 \xdef#1{{\rm(\number-\equanumber)}}%
     \gl@bal\advance\equanumber by -1
  \else \gl@bal\advance\equanumber by 1
     \ifx\chapterlabel\rel@x \def\d@t{}\else \def\d@t{.}\fi
    \xdef#1{{\rm(\chapterlabel\d@t\number\equanumber)}}\fi #1}}
\def\eqinsert#1{\noalign{\dimen@=\prevdepth \nointerlineskip
   \setbox0=\hbox to\displaywidth{\hfil #1}
   \vbox to 0pt{\kern 0.5\baselineskip\hbox{$\!\box0\!$}\vss}
   \prevdepth=\dimen@}}
%

%
%
\def\GENITEM#1;#2{\par \hangafter=0 \hangindent=#1
    \Textindent{$ #2 $}\ignorespaces}
\outer\def\newitem#1=#2;{\gdef#1{\GENITEM #2;}}

\newdimen\itemsize                \itemsize=30pt
\newitem\item=1\itemsize;
\newitem\sitem=1.75\itemsize;     
\newitem\ssitem=2.5\itemsize;     
\outer\def\newlist#1=#2&#3&#4;{\toks0={#2}\toks1={#3}%
   \count255=\escapechar \escapechar=-1
   \alloc@0\list\countdef\insc@unt\listcount     \listcount=0
   \edef#1{\par
      \countdef\listcount=\the\allocationnumber
      \advance\listcount by 1
      \hangafter=0 \hangindent=#4
      \Textindent{\the\toks0{\listcount}\the\toks1}}
   \expandafter\expandafter\expandafter
    \edef\c@t#1{begin}{\par
      \countdef\listcount=\the\allocationnumber \listcount=1
      \hangafter=0 \hangindent=#4
      \Textindent{\the\toks0{\listcount}\the\toks1}}
   \expandafter\expandafter\expandafter
    \edef\c@t#1{con}{\par \hangafter=0 \hangindent=#4 \noindent}
   \escapechar=\count255}
\def\c@t#1#2{\csname\string#1#2\endcsname}
\newlist\point=\Number&.&1.0\itemsize;
\newlist\subpoint=(\alphabetic&)&1.75\itemsize;
\newlist\subsubpoint=(\roman&)&2.5\itemsize;
%

%
%
%
%
\newcount\referencecount     \referencecount=0
\newcount\lastrefsbegincount \lastrefsbegincount=0
\newif\ifreferenceopen       \newwrite\referencewrite
\newdimen\refindent          \refindent=30pt
\def\normalrefmark#1{\attach{\scriptscriptstyle [ #1 ] }}
\let\PRrefmark=\attach
\def\NPrefmark#1{\step@ver{{\;[#1]}}}
\def\refmark#1{\rel@x\ifPhysRev\PRrefmark{#1}\else\normalrefmark{#1}\fi}
\def\refend@{\refmark{\number\referencecount}}
\def\refend{\refend@{}\space }
\def\refsend{\refmark{\count255=\referencecount
   \advance\count255 by-\lastrefsbegincount
   \ifcase\count255 \number\referencecount
   \or \number\lastrefsbegincount,\number\referencecount
   \else \number\lastrefsbegincount-\number\referencecount \fi}\space }
\def\REFNUM#1{\rel@x \gl@bal\advance\referencecount by 1
    \xdef#1{\the\referencecount }}
\def\Refnum#1{\REFNUM #1\refend@ } 
\def\REF#1{\REFNUM #1\R@FWRITE\ignorespaces}
\def\Ref#1{\Refnum #1\REFWRITE }
\def\ref{\Ref\?}
\def\REFS#1{\REFNUM #1\gl@bal\lastrefsbegincount=\referencecount
    \REFWRITE }

\def\r@fitem#1{\par \hangafter=0 \hangindent=\refindent \Textindent{#1}}
\def\refitem#1{\r@fitem{#1.}}
\def\NPrefitem#1{\r@fitem{[#1]}}
\def\NPrefs{\let\refmark=\NPrefmark \let\refitem=\NPrefitem}
\def\REFWRITE{\R@FWRITE\rel@x }
\def\R@FWRITE#1{\ifreferenceopen \else \gl@bal\referenceopentrue
     \immediate\openout\referencewrite=\jobname.refs
     \toks@={\begingroup \refoutspecials \catcode`\^^M=10 }%
     \immediate\write\referencewrite{\the\toks@}\fi
    \immediate\write\referencewrite{\noexpand\refitem %
                                    {\the\referencecount}}%
    \p@rse@ndwrite \referencewrite #1}
\begingroup
 \catcode`\^^M=\active \let^^M=\relax %
 \gdef\p@rse@ndwrite#1#2{\begingroup \catcode`\^^M=12 \newlinechar=`\^^M%
         \chardef\rw@write=#1\sc@nlines#2}%
 \gdef\sc@nlines#1#2{\sc@n@line \g@rbage #2^^M\endsc@n \endgroup #1}%
 \gdef\sc@n@line#1^^M{\expandafter\toks@\expandafter{\deg@rbage #1}%
         \immediate\write\rw@write{\the\toks@}%
         \futurelet\n@xt \sc@ntest }%
\endgroup
\def\sc@ntest{\ifx\n@xt\endsc@n \let\n@xt=\rel@x
       \else \let\n@xt=\sc@n@notherline \fi \n@xt }
\def\sc@n@notherline{\sc@n@line \g@rbage }
\def\deg@rbage#1{}
\let\g@rbage=\relax    \let\endsc@n=\relax
\def\refout{\par\penalty-400\vskip\chapterskip
   \spacecheck\referenceminspace
   \ifreferenceopen \Closeout\referencewrite \referenceopenfalse \fi
   \line{\fourteenrm\hfil REFERENCES\hfil}\vskip\headskip
   \input \jobname.refs
   }
\def\refoutspecials{\sfcode`\.=1000 \interlinepenalty=1000
         \rightskip=\z@ plus 1em minus \z@ }
\def\Closeout#1{\toks0={\par\endgroup}\immediate\write#1{\the\toks0}%
   \immediate\closeout#1}
%
%
\newcount\figurecount     \figurecount=0
\newcount\tablecount      \tablecount=0
\newif\iffigureopen       \newwrite\figurewrite
\newif\iftableopen        \newwrite\tablewrite
\def\FIGNUM#1{\rel@x \gl@bal\advance\figurecount by 1
    \xdef#1{\the\figurecount}}
\def\FIGURE#1{\FIGNUM #1\F@GWRITE\ignorespaces }

\def\figitem#1{\r@fitem{#1)}}
\def\FIGWRITE{\F@GWRITE\rel@x }
\def\TABNUM#1{\rel@x \gl@bal\advance\tablecount by 1
    \xdef#1{\the\tablecount}}
\def\TABLE#1{\TABNUM #1\T@BWRITE\ignorespaces }

\def\tabitem#1{\r@fitem{#1:}}
\def\TABWRITE{\T@BWRITE\rel@x }
\def\F@GWRITE#1{\iffigureopen \else \gl@bal\figureopentrue
     \immediate\openout\figurewrite=\jobname.figs
     \toks@={\begingroup \catcode`\^^M=10 }%
     \immediate\write\figurewrite{\the\toks@}\fi
    \immediate\write\figurewrite{\noexpand\figitem %
                                 {\the\figurecount}}%
    \p@rse@ndwrite \figurewrite #1}
\def\T@BWRITE#1{\iftableopen \else \gl@bal\tableopentrue
     \immediate\openout\tablewrite=\jobname.tabs
     \toks@={\begingroup \catcode`\^^M=10 }%
     \immediate\write\tablewrite{\the\toks@}\fi
    \immediate\write\tablewrite{\noexpand\tabitem %
                                 {\the\tablecount}}%
    \p@rse@ndwrite \tablewrite #1}
\def\figout{\par\penalty-400
   \vskip\chapterskip\spacecheck\referenceminspace
   \iffigureopen \Closeout\figurewrite \figureopenfalse \fi
   \line{\fourteenrm\hfil FIGURE CAPTIONS\hfil}\vskip\headskip
   \input \jobname.figs
   }
\def\tabout{\par\penalty-400
   \vskip\chapterskip\spacecheck\referenceminspace
   \iftableopen \Closeout\tablewrite \tableopenfalse \fi
   \line{\fourteenrm\hfil TABLE CAPTIONS\hfil}\vskip\headskip
   \input \jobname.tabs
   }
%
%
%
\newbox\picturebox
\def\p@cht{\ht\picturebox }
\def\p@cwd{\wd\picturebox }
\def\p@cdp{\dp\picturebox }
\newdimen\xshift
\newdimen\yshift
\newdimen\captionwidth
\newskip\captionskip
\captionskip=15pt plus 5pt minus 3pt
\def\fullwidth{\captionwidth=\hsize }
\newtoks\Caption
\newif\ifcaptioned
\newif\ifselfcaptioned
\def\caption{\captionedtrue \Caption }
\newcount\linesabove
\newif\iffileexists
\newtoks\picfilename
\def\fil@#1 {\fileexiststrue \picfilename={#1}}
\def\file#1{\if=#1\let\n@xt=\fil@ \else \def\n@xt{\fil@ #1}\fi \n@xt }
\def\pl@t{\begingroup \pr@tect
    \setbox\picturebox=\hbox{}\fileexistsfalse
    \let\height=\p@cht \let\width=\p@cwd \let\depth=\p@cdp
    \xshift=\z@ \yshift=\z@ \captionwidth=\z@
    \Caption={}\captionedfalse
    \linesabove =0 \picturedefault }
\def\plot{\pl@t \selfcaptionedfalse }
\def\Picture#1{\gl@bal\advance\figurecount by 1
    \xdef#1{\the\figurecount}\pl@t \selfcaptionedtrue }

\def\s@vepicture{\iffileexists \parsefilename \redopicturebox \fi
   \ifdim\captionwidth>\z@ \else \captionwidth=\p@cwd \fi
   \xdef\lastpicture{\iffileexists
        \setbox0=\hbox{\raise\the\yshift \vbox{%
              \moveright\the\xshift\hbox{\picturedefinition}}}%
        \else \setbox0=\hbox{}\fi
         \ht0=\the\p@cht \wd0=\the\p@cwd \dp0=\the\p@cdp
         \vbox{\hsize=\the\captionwidth \line{\hss\box0 \hss }%
              \ifcaptioned \vskip\the\captionskip \noexpand\Tenpoint
                \ifselfcaptioned Figure~\the\figurecount.\enspace \fi
                \the\Caption \fi }}%
    \endgroup }
\let\endpicture=\s@vepicture
\def\savepicture#1{\s@vepicture \global\let#1=\lastpicture }
\def\displaypicture{\fullwidth \s@vepicture $$\lastpicture $${}}
\def\toppicture{\fullwidth \s@vepicture \topinsert
    \lastpicture \medskip \endinsert }
\def\midpicture{\fullwidth \s@vepicture \midinsert
    \lastpicture \endinsert }
%
%
\def\leftpicture{\pres@tpicture
    \dimen@i=\hsize \advance\dimen@i by -\dimen@ii
    \setbox\picturebox=\hbox to \hsize {\box0 \hss }%
    \wr@paround }
\def\rightpicture{\pres@tpicture
    \dimen@i=\z@
    \setbox\picturebox=\hbox to \hsize {\hss \box0 }%
    \wr@paround }
\def\pres@tpicture{\gl@bal\linesabove=\linesabove
    \s@vepicture \setbox\picturebox=\vbox{
         \kern \linesabove\baselineskip \kern 0.3\baselineskip
         \lastpicture \kern 0.3\baselineskip }%
    \dimen@=\p@cht \dimen@i=\dimen@
    \advance\dimen@i by \pagetotal
    \par \ifdim\dimen@i>\pagegoal \vfil\break \fi
    \dimen@ii=\hsize
    \advance\dimen@ii by -\parindent \advance\dimen@ii by -\p@cwd
    \setbox0=\vbox to\z@{\kern-\baselineskip \unvbox\picturebox \vss }}
\def\wr@paround{\Caption={}\count255=1
    \loop \ifnum \linesabove >0
         \advance\linesabove by -1 \advance\count255 by 1
         \advance\dimen@ by -\baselineskip
         \expandafter\Caption \expandafter{\the\Caption \z@ \hsize }%
      \repeat
    \loop \ifdim \dimen@ >\z@
         \advance\count255 by 1 \advance\dimen@ by -\baselineskip
         \expandafter\Caption \expandafter{%
             \the\Caption \dimen@i \dimen@ii }%
      \repeat
    \edef\n@xt{\parshape=\the\count255 \the\Caption \z@ \hsize }%
    \par\noindent \n@xt \strut \vadjust{\box\picturebox }}
\let\picturedefault=\relax
\let\parsefilename=\relax
\def\redopicturebox{\let\picturedefinition=\rel@x
   \errhelp=\disabledpictures
   \errmessage{This version of TeX cannot handle pictures.  Sorry.}}
\newhelp\disabledpictures
     {You will get a blank box in place of your picture.}
%
%
%
%
%
%
%
%
%
%
\def\FRONTPAGE{\ifvoid255\else\vfill\penalty-20000\fi
   \gl@bal\pagenumber=1     \gl@bal\chapternumber=0
   \gl@bal\equanumber=0     \gl@bal\sectionnumber=0
   \gl@bal\referencecount=0 \gl@bal\figurecount=0
   \gl@bal\tablecount=0     \gl@bal\frontpagetrue
   \gl@bal\lastf@@t=0       \gl@bal\footsymbolcount=0}

\def\papers{\papersize\headline=\paperheadline\footline=\paperfootline}
\def\papersize{
   \advance\hoffset by\HOFFSET \advance\voffset by\VOFFSET
   \pagebottomfiller=0pc
   \skip\footins=\bigskipamount \normalspace }
\papers  
%
%
\newskip\lettertopskip       \lettertopskip=20pt plus 50pt
\newskip\letterbottomskip    \letterbottomskip=\z@ plus 100pt
\newskip\signatureskip       \signatureskip=40pt plus 3pt
\def\lettersize{\hsize=6.5in \vsize=8.5in \hoffset=0in \voffset=0.5in
   \advance\hoffset by\HOFFSET \advance\voffset by\VOFFSET
   \pagebottomfiller=\letterbottomskip
   \skip\footins=\smallskipamount \multiply\skip\footins by 3
   \singlespace }
\def\MEMO{\lettersize \headline=\letterheadline \footline={\hfil }%
   \let\rule=\memorule \FRONTPAGE \memohead }

\def\memodate{\afterassignment\MEMO \date }
\def\memit@m#1{\smallskip \hangafter=0 \hangindent=1in
    \Textindent{\caps #1}}
\def\subject{\memit@m{Subject:}}
\def\topic{\memit@m{Topic:}}
\def\from{\memit@m{From:}}
\def\memorule{\medskip\hrule height 1pt\bigskip}  
\def\memohead{\centerline{\fourteenrm MEMORANDUM}}
\newwrite\labelswrite
\newtoks\rw@toks
\def\letters{\lettersize
   \headline=\letterheadline \footline=\letterfootline
   \immediate\openout\labelswrite=\jobname.lab}

\let\letterhead=\rel@x
\def\addressee#1{\medskip\line{\hskip 0.75\hsize plus\z@ minus 0.25\hsize
                               \the\date \hfil }%
   \vskip \lettertopskip
   \ialign to\hsize{\strut ##\hfil\tabskip 0pt plus \hsize \crcr #1\crcr}
   \writelabel{#1}\medskip \noindent\hskip -\spaceskip \ignorespaces }
\def\rwl@begin#1\cr{\rw@toks={#1\crcr}\rel@x
   \immediate\write\labelswrite{\the\rw@toks}\futurelet\n@xt\rwl@next}
\def\rwl@next{\ifx\n@xt\rwl@end \let\n@xt=\rel@x
      \else \let\n@xt=\rwl@begin \fi \n@xt}
\let\rwl@end=\rel@x
\def\writelabel#1{\immediate\write\labelswrite{\noexpand\labelbegin}
     \rwl@begin #1\cr\rwl@end
     \immediate\write\labelswrite{\noexpand\labelend}}
\newtoks\FromAddress         \FromAddress={}
\newtoks\sendername          \sendername={}
\newbox\FromLabelBox
\newdimen\labelwidth          \labelwidth=6in
\def\makelabels{\afterassignment\Makelabels \sendersname=}
\def\Makelabels{\FRONTPAGE \letterinfo={\hfil } \MakeFromBox
     \immediate\closeout\labelswrite  \input \jobname.lab\vfil\eject}
\let\labelend=\rel@x
\def\labelbegin#1\labelend{\setbox0=\vbox{\ialign{##\hfil\cr #1\crcr}}
     \MakeALabel }
\def\MakeFromBox{\gl@bal\setbox\FromLabelBox=\vbox{\Tenpoint
     \ialign{##\hfil\cr \the\sendername \the\FromAddress \crcr }}}
\def\MakeALabel{\vskip 1pt \hbox{\vrule \vbox{
        \hsize=\labelwidth \hrule\bigskip
        \leftline{\hskip 1\parindent \copy\FromLabelBox}\bigskip
        \centerline{\hfil \box0 } \bigskip \hrule
        }\vrule } \vskip 1pt plus 1fil }
\def\signed#1{\par \nobreak \bigskip \dt@pfalse \begingroup
  \everycr={\noalign{\nobreak
            \ifdt@p\vskip\signatureskip\gl@bal\dt@pfalse\fi }}%
  \tabskip=0.5\hsize plus \z@ minus 0.5\hsize
  \halign to\hsize {\strut ##\hfil\tabskip=\z@ plus 1fil minus \z@\crcr
          \noalign{\gl@bal\dt@ptrue}#1\crcr }%
  \endgroup \bigskip }
\newbox\letterb@x
\def\lettertext{\par \vskip\parskip \unvcopy\letterb@x \par }
\def\multiletter{\setbox\letterb@x=\vbox\bgroup
      \everypar{\vrule height 1\baselineskip depth 0pt width 0pt }
      \singlespace \topskip=\baselineskip }
\def\letterend{\par\egroup}
%
%
%
\newskip\frontpageskip
\newtoks\Pubnum   
\newtoks\Pubtype  \let\pubtype=\Pubtype
\newif\ifp@bblock  \p@bblocktrue
\def\PH@SR@V{\doubl@true \baselineskip=24.1pt plus 0.2pt minus 0.1pt
             \parskip= 3pt plus 2pt minus 1pt }
\def\PHYSREV{\papers\PhysRevtrue\PH@SR@V}

\def\titlepage{\FRONTPAGE\papers\ifPhysRev\PH@SR@V\fi
   \ifp@bblock\p@bblock \else\hrule height\z@ \rel@x \fi }
\def\nopubblock{\p@bblockfalse}
\def\endpage{\vfil\break}
\frontpageskip=12pt plus .5fil minus 2pt
\Pubtype={}
\Pubnum={}
\def\p@bblock{\begingroup \tabskip=\hsize minus \hsize
   \baselineskip=1.5\ht\strutbox \topspace-2\baselineskip
   \halign to\hsize{\strut ##\hfil\tabskip=0pt\crcr
       \the\Pubnum\crcr\the\date\crcr\the\pubtype\crcr}\endgroup}
\def\title#1{\vskip\frontpageskip \titlestyle{#1} \vskip\headskip }
\def\author#1{\vskip\frontpageskip\titlestyle{\twelvecp #1}\nobreak}

\def\address#1{\par\kern 5pt\titlestyle{\twelvepoint\it #1}}
\def\andaddress{\par\kern 5pt \centerline{\sl and} \address}

\def\abstract{\par\dimen@=\prevdepth \hrule height\z@ \prevdepth=\dimen@
   \vskip\frontpageskip\centerline{\fourteenrm ABSTRACT}\vskip\headskip }

%
%
%

\def\\{\rel@x \ifmmode \backslash \else {\tt\char`\\}\fi }
\def\sequentialequations{\rel@x \if\equanumber<0 \else
  \gl@bal\equanumber=-\equanumber \gl@bal\advance\equanumber by -1 \fi }
\def\journal#1&#2(#3){\begingroup \let\journal=\dummyj@urnal
    \unskip, \sl #1\unskip~\bf\ignorespaces #2\rm
    (\afterassignment\j@ur \count255=#3), \endgroup\ignorespaces }
\def\j@ur{\ifnum\count255<100 \advance\count255 by 1900 \fi
          \number\count255 }
\def\dummyj@urnal{%
    \toks@={Reference foul up: nested \journal macros}%
    \errhelp={Your forgot & or ( ) after the last \journal}%
    \errmessage{\the\toks@ }}

\def\topspace{\hrule height 0pt depth 0pt \vskip}

\def\Buildrel#1\under#2{\mathrel{\mathop{#2}\limits_{#1}}}
\def\becomes#1{\mathchoice{\becomes@\scriptstyle{#1}}
   {\becomes@\scriptstyle{#1}} {\becomes@\scriptscriptstyle{#1}}
   {\becomes@\scriptscriptstyle{#1}}}
\def\becomes@#1#2{\mathrel{\setbox0=\hbox{$\m@th #1{\,#2\,}$}%
        \mathop{\hbox to \wd0 {\rightarrowfill}}\limits_{#2}}}
\def\Tr{\mathop{\rm Tr}\nolimits}

\let\int=\intop         
\def\lsim{\mathrel{\mathpalette\@versim<}}
\def\gsim{\mathrel{\mathpalette\@versim>}}
\def\@versim#1#2{\vcenter{\offinterlineskip
        \ialign{$\m@th#1\hfil##\hfil$\crcr#2\crcr\sim\crcr } }}
\def\big#1{{\hbox{$\left#1\vbox to 0.85\b@gheight{}\right.\n@space$}}}
\def\Big#1{{\hbox{$\left#1\vbox to 1.15\b@gheight{}\right.\n@space$}}}
\def\bigg#1{{\hbox{$\left#1\vbox to 1.45\b@gheight{}\right.\n@space$}}}
\def\Bigg#1{{\hbox{$\left#1\vbox to 1.75\b@gheight{}\right.\n@space$}}}
\def\){\mskip 2mu\nobreak }
%
%
%
\let\sec@nt=\sec
\def\sec{\rel@x\ifmmode\let\n@xt=\sec@nt\else\let\n@xt\section\fi\n@xt}
\def\obsolete#1{\message{Macro \string #1 is obsolete.}}
\def\firstsec#1{\obsolete\firstsec \section{#1}}
\def\firstsubsec#1{\obsolete\firstsubsec \subsection{#1}}
\def\thispage#1{\obsolete\thispage \gl@bal\pagenumber=#1\frontpagefalse}
\def\thischapter#1{\obsolete\thischapter \gl@bal\chapternumber=#1}
\def\splitout{\obsolete\splitout\rel@x}
\def\prop{\obsolete\prop \propto }
\def\nextequation#1{\obsolete\nextequation \gl@bal\equanumber=#1
   \ifnum\the\equanumber>0 \gl@bal\advance\equanumber by 1 \fi}
\def\BOXITEM{\afterassigment\B@XITEM\setbox0=}
\def\B@XITEM{\par\hangindent\wd0 \noindent\box0 }
%
%
%
\def\phyzzx{PHY\setbox0=\hbox{Z}\copy0 \kern-0.5\wd0 \box0 X}
        
\everyjob{\xdef\today{\monthname~\number\day, \number\year}
        \input myphyx.tex }
\message{ by V.K.}
%
\catcode`\@=12 
%

\twelvepoint
\baselineskip=24truept

\gdef\journal #1, #2, #3, 1#4#5#6{
      {\sl #1~}{\bf #2}, #3 (1#4#5#6)}
\def\SSP{\journal Sol. Stat. Phys., }
\def\PRB{\journal Phys. Rev. B, }
\def\PL{\journal Phys. Lett., }
\def\PRD{\journal Phys. Rev. D, }
\def\JPCS{\journal Phys. Chem. Sol., }
\def\PR{\journal Phys. Rev., }
\def\PRL{\journal Phys. Rev. Lett., }
\def\JPAMG{\journal J. Phys. A: Math. Gen., }
\def\NP{\journal Nucl. Phys., }
\def\Tr{\rm Tr}

\line{\hfill UdeM-LPN-TH-179}
\titlepage
\title{Effective action of a 2+1 dimensional system of nonrelativistic fermions
in the presence of a uniform magnetic field: dissipation effects}
\author{S. Sakhi }
\centerline{{\sl Laboratoire de physique Nucl\'eaire}}
\centerline{{\sl Universit\'e de Montr\'eal}}
\centerline{{\sl C.~P.~6128, Succ.~A,}}
\centerline{{\sl Montr\'eal, Qu\'ebec, Canada H3C 3J7}}
\abstract{The effective action of nonrelativistic fermions in 2+1 dimensions
is analyzed at finite temperature and chemical potential in the presence of a
uniform magnetic field perpendicular to the plane. The method used is
 a generalization of the derivative expansion technique. The induced
Chern-Simons
term is computed and shown to exhibit the Hall quantization. Effects of
 dissipation due to collisions are also analyzed.}

\endpage

\REF\PraGir{R.~Prange and S.~Girvin, {\sl The quantum Hall effect} (Springer
-Verlag), New York, 1987.
}

\REF\Lau{R.~B.~Laughlin, \PRL 50, 1395, 1983.
}

\REF\FetHanLau{A.~Fetter, C.~Hanna and R.~Laughlin, \PRB 39, 9679, 1989.
}

\REF\Wilc{F.~Wilczek, \PRL 48, 1144, 1982.
}

\REF\Col{S.~Coleman, {\sl Aspects of Symmetry}, (Cambridge University Press),
1985, pp. 136-144; Also see S.~Coleman and E.~Weinberg, \PRD 7, 1888,
1973.
}

\REF\NamJoL{Y.~Nambu and G.~Jona-Lasinio, \PR 122, 345, 1961; \PR 124,
246, 1961.
}

\REF\Schakel{A.M.~J.~Shackel, {\sl On broken symmetries in Fermi systems},
University of Amsterdam thesis, 1989.
}

\REF\Wein{S.~Weinberg, UTTG-18-93.
}

\REF\AitFra{I.~J.~R.~Aitchison and C.~M.~Fraser, \PL 146B, 63, 1984.
}

\REF\Ait{I.~J.~R.~Aitchison and C.~M.~Fraser, \PRD 31, 2605, 1985.
}

\REF\DasKha{A.~Das and A.~Kharev, \PRD 36, 2591, 1987.
}

\REF\BabDasPan{K.~S.~Babu, A.~Das and P.~Panigrahi, \PRD 36, 3725, 1987.
}

\REF\MacPanSak{R.~MacKenzie, P.~K.~Panigrahi and S.~Sakhi, \PRB 48, Vol.~07,
 3892,
1993.
}

\REF\Kub{R.~Kubo, S.~J.~Miyake and N.~Hashitsume, \SSP 17, 269, 1965.
}

\REF\SakVas{S.~Sakhi and P.~Vasilopoulos, work in progress.
}

\REF\RANSALSTR{S.~Randjabar-Daemi, A.~Salam and J.~Strathdee, \NP B340,
403, 1990.
}

\REF\PanRaySak{P.~Panigrahi, R.~Ray and B.~Sakita, \PRB 42, 4036, 1990.
}

\REF\Wel{H.~A.~Weldon, \PRD 26, 10, 1982.
}

\REF\Kap{J.~Kapusta, {\sl Finite temperature field theory}, Cambridge
University press,
(1989).
}

\REF\Arg{P.~N.~Argyres,\JPCS 4, 19, 1957.
}

\REF\EncSalFan{R.~C.~Enck, A.~S.~Saleh and H.~Y.~Fan,\PR 182, 790, 1962
}

\REF\PINNOZ{D.~Pines and P.~Nozieres, {\sl The theory of quantum liquids},
Vol.1, pp.183 and pp.138, 1966.
}

\REF\Frad{E.~Fradkin, {\sl Field theories of condensed matter systems},
Frontiers in physics; Vol.82, pp.333, 1991.
}

\REF\Mah{Gerald D.~Mahan, {\sl Many-particle physics}, Plenum Press, 1990.
}

\REF\KriLatNaf{I.~V.~Krive, S.~M.~Latinsky and S.~A.~Naftulin, \JPAMG 25,
L921, 1992.
}

\REF\MER{U.~Merkt in {\sl the physics of the two dimensional electron
gas}, edited by J.T.Devreese and F.M. Peeters, NATO ASI series B: Physics,
Vol.157, pp.293.
}

In many situations in condensed matter physics one deals with systems of
fermions in an electromagnetic field which has both dynamical and
background parts.  The fractional quantum Hall effect (FQHE) [\PraGir] is one
of
these interesting systems, where the electrons are submitted to a high
magnetic field perpendicular to the plane of confinement and interacting
through a strong Coulomb interaction. In this particular system,
correlation effects give rise to the Hall plateaus at fractional values of
the filling factor and are expected to cause Wigner crystallization at
lower densities [\Lau]. Another interesting system, believed to be relevant to
high
$T_c$ superconductors [\FetHanLau], is the anyons gas in 2+1 dimensions
[\Wilc].
 The particles
in this system can be viewed as electrons carrying flux tubes; this leads to
a statistical magnetic field proportional to the anyon density.

A useful tool in studying the low energy excitations in such systems is the
 effective action formalism of quantum field theory. This is simply defined
as the sum of all connected one-particle-irreducible vacuum diagrams in
the presence of any background field [\Col,\NamJoL]. It has been successfully
applied in the classic case of BCS superconductivity [\Schakel,\Wein],
where
 all the
properties of the superconductor including the gap field, free energy, and
so on, can be derived.

Formally the effective action is achieved by integrating out the fermions,
which results in an expression involving a fermionic determinant. One then has
 to concoct
a method of evaluating this determinant. The derivative expansion
technique [\AitFra] is one such method.
It has been extremely useful in the calculation of anomaly induced
vortices in 3+1 dimensions [\Ait], proving the bosonization in 1+1
dimensions [\DasKha], showing the origin of Chern-Simons term in 2+1 dimensions
[\BabDasPan]. Recently, combined with the large-N expansion
technique, this method has been applied successfully to a 2+1 dimensional
system of charged fermions
 interacting through a four-Fermi term and exhibiting superconductivity
via the Kosterlitz-Thouless mechanism [\MacPanSak]. However, so far, only plane
wave states have been used
as the underlying single particle states in this technique.
In this paper, we propose a
generalization of this technique to the case where the single particle
states are localized, such as Landau states in the presence of a uniform
magnetic field.

To illustrate the technique, we apply it to a system of nonrelativistic
fermions in 2+1 dimensions subjected to a uniform magnetic field
perpendicular to the plane and interacting with a fluctuating gauge field
whose dynamics can be described by the usual Maxwell term, or eventually a
Chern-Simons term. This system, in connection with anyonic
superconductivity, has been recently analysed in reference [\RANSALSTR]
and in reference [\PanRaySak] where a different
approach based on the inhomogeneity derivative expansion technique
has been used. Unlike that one, our technique gives the higher derivative
terms in a straightforward way and in the form of some known polynomials.

The calculation will be carried out at finite temperature and finite
chemical potential.
To obtain information on the influence of the electron gas on the dynamics
of the photon, it is useful to integrate over the electron degree of
freedom, obtaining the effective action for the gauge field. Here we show that
in
this process a
Chern-Simons
term emerges with a coefficient depending on the magnetic field and the
 temperature,
in agreement with the result derived in [\PanRaySak] and [\RANSALSTR]. Finally,
the full gauge
 field propagator
at finite temperature is constructed and is shown to be tranverse in
 accordance with gauge
invariance.

The action at finite temperature describing the fermions coupled to the gauge
 field is written as
$$
S=\int_0^{\beta}d\tau \int d^2x\, \psi^\dagger \bigl(ip_\tau +\mu -
 \epsilon(-i\nabla -e{\bf
A}-e{\bf a})+iea_\tau\bigr)\psi+S_{\rm kin}[a],
\eqno(1)$$
where ${\bf A}$ is the background field and the last term is the kinetic
term of the fluctuating gauge field ${\bf a}$.~$\mu$~is the
chemical potential of the system, and $\beta=1/T$ where $T$ is
 the absolute temperature of the
system.  We are assuming a parabolic dispersion for the fermions:~$\epsilon
({\bf k})={\bf k}^2/2m$.

The effective action $S_{\rm eff}$ of this system is a functional of
$a_\mu$ and is expressed as a path integral over the fermionic variables
$$ \int D\psi^\dagger D\psi\,\,e^S=e^{-S_{\rm eff}},
\eqno(2)$$
where
$$
S_{\rm eff}=-\Tr\, \ln\big\{ ip_\tau +\mu -
\epsilon(-i\nabla -e{\bf A}-e{\bf a})+iea_\tau\big\}+S_{\rm kin}[a].
\eqno(3)$$

Varying twice with respect to the gauge field ${\bf a_\mu}$, we get the
different
 components of
the polarization tensor $\Pi_{\mu\nu}$.  To illustrate the technique, we
present the
 derivation of
one of these components, $\Pi_{ij}$, in some detail. The other components
are computed in a similar way. For $\Pi_{ij}$, $a_\tau$ can be discarded
 since no derivative with respect to it is taken.

First one expands the logarithm inside the trace as follows
$$
\eqalign{S_{\rm eff}&=-\Tr\,\ln\left( ip_\tau +\mu -{\hat H_0}\right)-Tr\left(
{1\over
ip_\tau +\mu -{\hat H_0}}\,\Bigl[\, {e\over 2m}\,({\bf \Pi\cdot a}+{\bf
a\cdot\Pi})-
{e^2\over
2m}{\bf a}^2\,\Bigr]\right) \cr
&+{1\over 2}\left({e\over 2m}\right)^2Tr\left({1\over
ip_\tau +\mu -{\hat H_0}}\,({\bf \Pi\cdot a}+{\bf a\cdot\Pi})\,{1\over
ip_\tau +\mu -{\hat H_0}}\,({\bf \Pi\cdot a}+{\bf a\cdot\Pi})\,\right)+\cdots
\cr}
\eqno(4)
$$
Now by varying twice with respect to $a$ and setting $ a=0$, we get
$$
\displaylines{\quad{\delta^2 S_{\rm eff}\over \delta a^i(x)\delta
a^j(y)}={e^2\over
m}\,\Tr\left( {1\over  ip_\tau +\mu -{\hat H_0}}\,\delta^3 ({\hat
R}-x)\,\delta^3 ({\hat
R}-y)\right)\hfill\cr
\hfill{}+{1\over 2}\left({e\over 2m}\right)^2\Tr\,{1\over
ip_\tau +\mu -{\hat H_0}}\,\left[{\hat\Pi}^i\,\delta^3 ({\hat R}-x)
             +\delta^3 ({\hat R}-x)\,{\hat\Pi}^i\right] \,\hfill\cr
\hfill{}\times{1\over ip_\tau +\mu -{\hat H_0}}\,\left[{\hat\Pi}^j\,
\delta^3 ({\hat R}-y)+\delta^3 ({\hat
            R}-y)\,{\hat\Pi}^j\right].\hfill\llap{(5)}\cr}
$$
In these equations ${\hat H_0}$ is the unperturbed hamiltonian giving rise, in
this
 case, to Landau
states ${|\zeta\rangle}\equiv {|\ell,X\rangle}$ [\Kub]. ${\hat\Pi}^i$ is the
 momentum operator in the
presence  of the gauge field ${\bf A}$ given by $p^i-eA^i$ and
 ${\hat R}$ stands for the operators
$({\hat \tau},{\hat {\bf r}})$.
The defining commutation relations between these operators are [\Kub]
$${\hat H_0}={1\over 2m}{\bf {\hat \Pi}}^2\qquad\hbox{with }\qquad
\lbrack {\hat \Pi}_x , {\hat \Pi}_y\rbrack = -{i\over \ell_B^2}\eqno(6)$$
The energy associated with the
states $|\ell,X\rangle$ is given by $E_\zeta=(\ell+{1\over 2})\omega_c$ where
 $\omega_c={eB\over m}$ is the
cyclotron frequency and each state has a degeneracy ${m\omega_c\over 2\pi}
\equiv{1\over2\pi\ell_B^2}$.
The guiding center coordinates are defined by
$$ {\hat X}={\hat x}-\ell_B^2{\hat \Pi_y},\qquad {\hat Y}={\hat y}+\ell_B^2
{\hat \Pi_x},\eqno(7a)$$
with the commutation relations
$$\eqalignno{\lbrack{\hat X}, {\hat \Pi_i}\rbrack&=\lbrack{\hat Y},
 {\hat \Pi_i}\rbrack
=0,\cr
\lbrack{\hat X}, {\hat Y}\rbrack&=i\ell_B^2.&(7b)\cr}$$
In the basis $|\ell,X\rangle$ we have
$${\hat X}|X\rangle=X|X\rangle\qquad\hbox{and}\qquad{\hat
Y}|X\rangle=-i\ell_B^2{\partial\over \partial X}|X\rangle. \eqno(7c)$$
And finally,
$$ \lbrack {\hat \tau} , {\hat p}_\tau
\rbrack = i. \eqno(8)$$

To evaluate the trace in equation $(5)$, we first use a spectral
 representation
 of the delta function:
$$\delta^3 ({\hat R} -x)=\int {d^3q\over ({2\pi})^3}\,\, e^{iq\cdot({\hat R}
-x)},
\eqno(9)$$
and by making use of the commutation relation $(8)$ to move ${\hat\tau}$
 to the left, we obtain in
Fourier space
$$\eqalignno{ \Pi^{ij}({\bf q},\omega)&={e^2\over m}{1\over\beta}\sum_n
\sum_\zeta
{1\over i\omega_n+\mu-E_\zeta}\cr
&+{e^2\over
4m^2}{1\over\beta}\sum_n\sum_{\zeta\zeta^\prime}{\langle\zeta|({\hat\Pi}^i
e^{i{\bf q}{\hat
{\bf r}}}+ e^{i{\bf q}{\hat{\bf
r}}}{\hat\Pi}^i)|\zeta^\prime\rangle\langle\zeta^\prime|({\hat\Pi}^j
e^{-i{\bf q}{\hat
{\bf r}}}+ e^{-i{\bf q}{\hat{\bf r}}}{\hat\Pi}^j)|\zeta\rangle\over
(i\omega_n+\mu-E_\zeta)(i\omega_n-i\omega +\mu-E_{\zeta^\prime})}.&(10)\cr}
$$
where we have a sum over the so called Matsubara frequencies $\omega_n=2\pi
 (n+{1\over 2})/\beta$ and
a sum over Landau states.~$\omega$ is an imaginary Bose frequency.
The first sum can be done easily
by contour integration (see appendix); the result is
$$
\eqalign{\Pi^{ij}({\bf q},\omega )&={e^2\over m}\sum_\zeta f_\zeta\,
 \delta^{ij}+{e^2\over 4m^2}\sum_{\zeta\zeta^\prime}{f_\zeta-f_{\zeta^\prime}
\over
E_\zeta-E_{\zeta^\prime}-i\omega}\langle\zeta|({\hat\Pi}^ie^{i{\bf q}{\hat
 {\bf r}}}+
e^{i{\bf q}{\hat{\bf
r}}}{\hat\Pi}^i)|\zeta^\prime\rangle\cr
&\qquad\times\langle\zeta^\prime|({\hat\Pi}^j
e^{-i{\bf q}{\hat {\bf
r}}}+ e^{-i{\bf q}{\hat{\bf r}}}{\hat\Pi}^j)|\zeta\rangle.\cr}\eqno(11)
$$
where $f_\zeta$ is the familiar Fermi-Dirac distribution function:
$$
f_\zeta={1\over 1+e^{\beta (E_\zeta -\mu)}}.
$$

Using the same procedure, we obtain the other components of the polarization
 tensor; these are
given by
$$
\eqalignno{\Pi^{i\tau}&={ie^2\over 2m}\sum_{\zeta\zeta^\prime}{f_\zeta-
f_{\zeta^\prime}\over
E_\zeta-E_{\zeta^\prime}-i\omega}\langle\zeta|({\hat\Pi}^ie^{i{\bf q}{\hat
 {\bf r}}}+
e^{i{\bf q}{\hat{\bf
r}}}{\hat\Pi}^i)|\zeta^\prime\rangle\langle\zeta^\prime|e^{-i{\bf q}{\hat {\bf
r}}}|\zeta\rangle&(12)\cr
\noalign{\hbox{and}}
\Pi^{\tau\tau}&=-e^2\sum_{\zeta\zeta^\prime}{f_\zeta-f_{\zeta^\prime}\over
E_\zeta-E_{\zeta^\prime}-i\omega}|\langle\zeta|e^{i{\bf q\cdot
r}}|\zeta^\prime\rangle|^2.&(13)\cr}
$$
It is worth mentioning that this last equation is related to the
density-density correlation function and that its analysis at
finite frequency $\omega$ gives information about
 so-called magnetoplasmon collective excitations. A detailed investigation of
these excitations along with other results will be reported elsewhere
[\SakVas].

Having these components, we can now construct the transverse polarization
tensor
 $\Pi^{\mu\nu}$.
Due to rotational invariance and current conservation [\Wel,\Kap],
it is expressed in terms of four tensors,
$$
\Pi^{\mu\nu}=\Pi_t\, T^{\mu\nu}+\Pi_l\, L^{\mu\nu}+\Pi_0\, E^{\mu\nu}+\Pi_p\,
 P^{\mu\nu},\eqno(14)
$$
which are defined as
$$
\eqalignno{T_{\mu\nu}&=\delta_\mu^i\delta_\nu^j({\bf
q^2}\delta_{ij}-q_iq_j),\cr
L_{\mu\nu}&=-q_\mu q_\nu +q^2 g_{\mu\nu}-T_{\mu\nu},\cr
E_{\mu\nu}&=\varepsilon_{\mu\nu\lambda}\,q^{\lambda},\cr
P_{\mu\nu}&=(\delta_\mu^i\delta_\nu^0+\delta_\mu^0\delta_\nu^i)\,
\varepsilon_{ki}\,q^k{\bf
q}^2\cr
&\qquad+\delta_\mu^i\delta_\nu^j(\varepsilon_{ik}\,q^j+\varepsilon_{jk}\,q^i)
q^kq^0.&(15)
\cr}$$
These tensors correspond respectively to the $B^2$, $E^2$, Chern-Simons and
 $B{\bf\nabla  E}$
terms in the effective action. Although the last term is irrelevant in the
infrared limit, it is included here because the algebra formed by the first
three tensors does not close without it.The coefficients
$\Pi_t,\,\Pi_l,\,\Pi_0,\,
\Pi_p$ are analytical
functions of ${\bf q}\,\hbox{and}\,\omega$; they are expressed as
$$
\eqalignno{\Pi_l&=-{1\over \bf q^2}\,\Pi^{00},\cr
\Pi_t&={1\over \bf q^2}\left(\Pi^{ii}+2{q_0^2\over \bf
q^2}\,\Pi^{00}\right),\cr
\Pi_0&={q_\lambda \varepsilon^{\mu\nu\lambda}\over q^2}\,\Pi_{\mu\nu},\cr
\Pi^{i0}&=q^i\,q^0\Pi_l+\varepsilon^{ij}q^j(\Pi_0+{\bf q}^2\Pi_p).&(16)\cr}
$$
In computing the different components of the polarization tensor, one makes
 use of the following
relations [\Arg,\EncSalFan]:
$$
\eqalignno{|\langle\zeta|e^{i{\bf q\cdot r}}|\zeta^\prime\rangle|^2&=
|\langle\ell,X|e^{i{\bf q\cdot r}}|\ell^\prime,X^\prime\rangle|^2\cr
&=|J_{\ell\ell^\prime}(u)|^2\delta_{X,X^\prime+\ell_B^2 q_y},&(17a)\cr}
$$
where $u=\ell_B^2(q_x^2+q_y^2)/2,$
$$|J_{\ell\ell^\prime}(u)|^2=\bigl(\ell!\,/\ell^\prime!\bigr)e^{-u}\,u^{
\ell^\prime-\ell}
\bigl[L_\ell^{\ell^\prime-\ell}(u)\bigr]^2,\eqno(17b)
$$
and $L_\ell^{\ell^\prime-\ell}$ is an associated Laguerre polynomial for
$\ell^\prime\geq\ell $ (for  $\ell^\prime\leq\ell $, interchange $\ell$ and
 $\ell^\prime$
in this formula).
 It is also helpful to use the following relation, which is easy to
establish:
$$
\Bigl\{{\hat\Pi}^k,e^{i\ell_B^2\varepsilon_{ij}q^i{\hat\Pi}^j}\Bigr\}=
\varepsilon^{nk}{2\over
i\ell_B^2} {\partial \over \partial
q^n }\,e^{i\ell_B^2\varepsilon_{ij}q^i{\hat\Pi}^j},\eqno(18) $$
here the curly braces denote the anticommutator.

In principle, using these relations, it is straightforward, although tedious,
to evaluate exactly
the different components of the polarization tensor. However, we shall give
 here only the long wavelength limit of these coefficients. At finite
temperature, it turns out that $\Pi_l$ has a pole when $q\to 0$ and
$\omega=0$ indicating that the limits $(q,\omega)\to 0$ and $T\to 0$ don't
commute [\PINNOZ].
Consequently the effective action has a nonlocal electric-electric field
term. This is reminiscent of the anomalous skin effect in metals where the
static transverse conductivity shows the same singular behavior
$\sigma(q,0)=C/q$
[\PINNOZ].
In the long wavelength limit we obtain
$$
\eqalignno{\Pi_l&={e^2m\omega_c\over 2\pi}{1\over
q^2}\left(\sum_{\ell=0}^\infty {\partial f\over\partial E_\zeta}\right)
-{e^2\over 2\pi\omega_c}\left(\sum_{\ell=0}^\infty f_\ell
\right)-{e^2\over 2\pi}\left(\sum_{\ell=0}^\infty {\partial f\over\partial
E_\zeta}(\ell+{1\over 2})\right),&(19a)\cr
\Pi_t&={e^2\over\pi m}\left(\sum_{\ell=0}^\infty (\ell +{1\over 2})f_\ell
\right)+{e^2\omega_c\over 2\pi m}\left(\sum_{\ell=0}^\infty {\partial
f\over\partial
E_\zeta}(\ell+{1\over 2})^2\right),&(19b)\cr
\Pi_0&={e^2\over 2\pi}\left(\sum_{\ell=0}^\infty f_\ell\right)+
{e^2\omega_c\over 2\pi}\left(\sum_{\ell=0}^\infty {\partial f\over\partial
E_\zeta}(\ell+{1\over 2})\right),&(19c)\cr
\Pi_p&=-{e^2\over \pi m\omega_c}\left(\sum_{\ell=0}^\infty (\ell +{1\over
2})f_\ell\right).&(19d)\cr}
$$
To gain more insight, we now consider the zero temperature limit when the
 Fermi level is pinned
in a gap between two
Landau levels
$$
E_N\,<\epsilon_F\,<\,E_{N+1}.
$$
In this case, we make the replacement
$$
\sum_{\ell=0}^\infty f_\ell \longrightarrow (N+1)\qquad\hbox{and}\qquad
\sum_{\ell=0}^\infty (\ell +{1\over 2})f_\ell \longrightarrow {(N+1)^2\over 2}
\qquad N=0,1,\ldots
$$
The coefficient of the Chern-Simons term is thus
$$
\Pi_0={e^2\over 2\pi}(N+1).\eqno(20)
$$
This coefficient is exactly the Hall
 conductivity
$\sigma_{yx}$ [\Frad], and hence $eq.(20)$ expresses the famous integral
quantum Hall effect
 discovered by Von
Klitzing in 1980.

Next, we consider the effect of dissipation due to collisions with impurities.
This is known to be of considerable
importance for the 2D gas, especially in connection with transport properties.
 This effect is accounted for by means
of a relaxation time $\tau$, which is usually  obtained by computing the
imaginary
part of the fermions self-energy in the presence of a scattering potential.
However, one has to do the calculation in a
self-consitent way because of polarization effects. Furthermore, this
relaxation
time, in general, depends on each Landau level.
Hereafter, we assume a single finite relaxation, and compute its
effect on the polarization tensor. The fermion propagator is then modified to
[\Schakel,\Mah]
$$
{1\over i\omega_n +\mu -E_\zeta -i{{\rm sgn}(\omega_n)\over 2\tau}}\eqno(21)
$$
which can also be written as
$$
\int_{-\infty}^{+\infty}\,dk_0\,\sigma (k_0)\,{1\over i\omega_n -k_0}\eqno(22a)
$$
where use has been made of the spectral function
$$
\sigma (k_0)={1\over 2\pi\tau}\,{1\over k_0^2+{1\over 4\tau^2}}.\eqno(22b)
$$
Consequently, in equations (11), (12) and (13) one has then to make the
following substitution
$$
\sum_{\zeta\zeta^\prime}{f_\zeta-f_{\zeta^\prime}\over
E_\zeta-E_{\zeta^\prime}-i\omega}\longrightarrow\sum_{\zeta\zeta^\prime}
\int_{-\infty}^{+\infty}dk_0\,\int_{-\infty}^{+\infty}dk_0^{\prime}\,
\sigma(k_0-E_\zeta)
\sigma(k_0^{\prime}-E_{\zeta^\prime})
\,{f(k_0)-f(k_0^\prime)\over k_0-k_0^\prime-i\omega}.\eqno(23)
$$
We find it here more interesting to determine the effect of a finite life
 time on the Chern-Simons
coefficient, since the quantization shown in equation (20) has a
 crucial consequence
on two dimensional electron gas. Using the substitution (23) in the expression
of $\Pi_0$ given in (16),
 we find
$$
\Pi_0(\omega)={e^2\omega_c^2\over 2\pi}\sum_{\ell=0}^{\infty}(\ell+1)\int_{-
\infty}^{+\infty}dk_0\,
\int_{-\infty}^{+\infty}dk_0^{\prime}\,\sigma(k_0-E_\ell)
\sigma(k_0^{\prime}-E_{\ell+1})
\,{f(k_0)-f(k_0^\prime)\over (k_0-k_0^\prime)^2+\omega^2}\eqno(24)
$$
Next, we perform the $k_0^{\prime}$ integration in the first term and the
 $k_0$ integration in the
second term of this equation, giving
$$
\Pi_0(\omega)={e^2\omega_c^2\over 2\pi}\sum_{\ell=0}^{\infty}\int_{-\infty}^{+
\infty}dk_0\,f(k_0)\,
\sigma(k_0-E_\ell)\Bigl\{(\ell+1)\,g_\omega(k_0-E_{\ell+1})-\ell\,g_\omega(k_0-
E_{\ell-1})\Bigr\},\eqno(25a)
$$
where we have introduced the function
$$
\eqalignno{g_\omega(x)&=\int_{-\infty}^{+\infty}dk_0\,{\sigma(k_0)\over (k_0+x)
^2+\omega^2}\cr
&=\Bigl(1+{1\over \omega\tau}\Bigr)\,{1\over x^2+(\omega+1/2\tau)^2)}.
&(25b)\cr}
$$
Now in a strong magnetic field we have $\omega_c\tau\gg 1$ and equation (25a)
 reduces to
$$
\Pi_0(\omega)={e^2\rho\over m}\,\omega_c\,g_\omega(\omega_c),\eqno(26)
$$
where $\rho$ is the particle density given by
$$
\rho={m\omega_c\over 2\pi}\sum_{\ell=0}^{\infty}\int_{-\infty}^{+\infty}dk_0\,
f(k_0)\,
\sigma(k_0-E_\ell),
$$
and the static Chern-Simons coefficient is
$$
\Pi_0(0)={e^2\rho\over m}\,{\omega_c\tau^2\over \tau^2\omega_c^2+1/4}\eqno(27)
$$
In the absence of dissipation ($\tau\to\infty$), this expression reduces to
that
given in equation (19c).
For the sake of comparison, we give here the same induced coefficient in
the case
of massive Dirac fermions with dissipation effects when the magnetic field is
absent [\KriLatNaf].  In this case, we obtain
$$
\Pi_0={1\over 2}{\partial\over\partial m}\left(\rho/m\right);\eqno(28a)
$$
for relativistic fermions the particles density is given by
$$
\rho={2m\over \beta}\sum_n\int\,{d^2p\over(2\pi)^2}\,{1\over (\omega_n+
{\rm sgn}(\omega_n)/2\tau-i\mu)^2+{\bf p}^2+m^2},\eqno(28b)
$$
from this, we get at $T=0$
$$
\Pi_0={1\over 4\pi^2}\, {|m|\over m}\left\{\arctan 2\tau(|m|+\mu)+\arctan 2\tau
(|m|-\mu)\right\}.\eqno(28c)
$$
In the absence of dissipation, this equation reproduces the by now familiar
Chern-Simons term $\Pi_0={\rm sgn}(m)/4\pi$.

In concluding we would like first to emphasize that this method is useful for
any
background field for which we know the exact eigenstates. In particular
equations (10), (11), (12) and (13) are exactly the same when the
background field has a constant magnetic field as well as a constant
electric field not too strong to prevent the destruction of Landau levels
quantization [\MER] and the particles creation, the only changement to be made
is in the eigenstates' energy
which is now shifted by the electric field and consequently the degeneracy
is lifted.
Another advantage of this technique is the exact form, in terms of Laguerre
plynomials,
obtained for the higher
derivative terms (or in Fourier space for any $({\bf q},\omega)$). As noted in
the maniscript, this is of particular interest for the magnetoplasmon
collective excitations, when one is concerned with higher values of ${\bf
q}$.
Application of this technique to the case of massive and massless Dirac
fermions in the presence of a uniform magnetic field is currently under
progress and will be reported elsewhere.

I am grateful to P.~Panigrahi for explaining to me their method in ref.~16. I
also thank R.~MacKenzie and M.~Paranjape for their encouragement and their
valuable comments. This work was supported in
part by the Natural Science and Engineering Research Council of Canada and the
fonds F.~C.~A.~R du Qu\'ebec.

\appendix

In this appendix, we evaluate the Matsubara sum appearing in equations
(10),(11),(12) and (13). We adopt the standard technique of contour
integration. The summation are of the form
$$
S=-{1\over \beta}\sum_{n=-\infty}^{n=+\infty}\,g(i\omega_n)
\eqno(A.1)
$$
where $\omega_n={2\pi\over \beta}(n+1/2)$.
We shall do an integral of the form
$$
I=lim_{R\to\infty}\int {dz\over 2\pi i}\,f(z)\,g(z)\eqno(A.2)
$$
where the contour is a large circle of radius R in the limit $R\to \infty$.
The function $f(z)$ is chosen to generate poles at the points $i\omega_n$.
The function which does this is
$$
f(z)={1\over 1+e^{\beta z}}
,$$
the residue at its poles is $(-1/\beta)$.
In the limit $R\to \infty$, the integral vanishes, $I=0$ so that the
summation is given by the residues
$$
S=\sum \quad\hbox{residue of}\quad f(z)g(z)\quad \hbox{at the poles
of}\quad g(z).\eqno(A.3)
$$
Applying this procedure we derive the following formula
$$
{1\over\beta}\sum_n {1\over i\omega_n+\mu-E_\zeta}=f_\zeta
\eqno(A.4)$$
and
$$
{1\over\beta}\sum_n {1\over (i\omega_n+\mu-E_\zeta)^2}={\partial
f_\zeta\over\partial E_\zeta}
\eqno(A.5)$$
and finaly
$$
{1\over\beta}\sum_n\sum_{\zeta\zeta^\prime}{F_{\zeta\zeta^\prime}(q)
\over(i\omega_n+\mu-E_\zeta)(i\omega_n+\mu-E_{\zeta^\prime}-i\omega)}=\sum_\zeta
{\partial f_\zeta\over\partial
E_\zeta}\,F_{\zeta\zeta}(q)\,\delta_{\omega=0}$$
$$+\sum_{\zeta\neq\zeta^\prime}
{f_\zeta-f_{\zeta^\prime}\over
E_\zeta-E_{\zeta^\prime}-i\omega}\,F_{\zeta\zeta^\prime}(q)
\eqno(A.6)$$
where we have used the identity
$$
f(E_\zeta-i\omega)=f(E_\zeta)
$$
because $\omega$, beeing a Bose frequency, is given by $2\pi s/\beta$ with
$s$
an integer.
\refout

\bye